\newcommand{\be}{\begin{equation}}\newcommand{\ee}{\end{equation}}
\newcommand{\bea}{\begin{eqnarray}}\newcommand{\eea}{\end{eqnarray}}
\newcommand{\nn}{\nonumber}\newcommand{\p}[1]{(\ref{#1})}
 \newcommand{\lb}[1]{\label{#1}}
\newcommand\q{\quad}\newcommand\qq{\qquad}

\newcommand\hT{\hat\Theta}

\newcommand\adb{{\alpha\dot{\beta}}}

\newcommand\bg{{\beta\gamma}}

\newcommand\da{{\dot{\alpha}}}

\newcommand\cV{{\cal V}}
\newcommand\cD{{\cal D}}

\newcommand\hcD{\hat{\cal D}}

\newcommand\s{\scriptscriptstyle}
\newcommand\A{{\s A}}
\newcommand\B{{\s B}}

\newcommand\R{{\s R}}

\newcommand{\pp}{{\s ++}}
\newcommand{\m}{{\s --}}

\newcommand{\Dp}{D^{\pp}}
\newcommand{\Dm}{D^{\m}}
\newcommand{\Vp}{V^{\pp}}
\newcommand{\Vm}{V^{\m}}

\newcommand{\DP}{D^+}

\newcommand{\bDP}{\bar{D}^+}

\documentclass[12pt]{article}

\begin{document}
\renewcommand{\thefootnote}{\fnsymbol{footnote}}
\begin{flushright}
{ hep-th/0109113 }
\end{flushright}
\vspace{2cm}

\begin{center}
{\large\bf   STATIC BPS-CONDITIONS
 IN N=2 HARMONIC SUPERSPACE\footnote{Talk at the 9th international
 conference on supersymmetry and unification of fundamental
 interactions, Dubna, Russia, June 11-17, 2001}}
\vspace{1cm} \\

{\bf B.M. Zupnik}\\
\vspace{0.5cm}

{\it Joint Institute for Nuclear Research, Bogoliubov Laboratory
of Theoretical Physics, Dubna, Moscow Region,
 141980, Russia. E-mail: zupnik@thsun1.jinr.ru }

\end{center}

\begin{abstract}
We analyze superfield representations of BPS-conditions for the
self-dual  static  solutions of $D=4, N=2$ supersymmetric Yang-Mills
theory.
\end{abstract}

The off-shell superfield constraints of the $N=2 $ super-Yang-Mills
(SYM) theory in the Minkowski $D=(3,1)$ space has been solved in the
framework of harmonic superspace (HSS) using the auxiliary  coordinates
of the coset space $S_2\sim SU_\A(2)/U_\A(1)$ where $SU_\A(2)$  is the
corresponding automorphism group \cite{GIK1}-\cite{Zu2}.

Harmonic variables connected with coset $SU_\R(2)/U_\R(1)$ of the
subgroup of the Euclidean rotation group have been used to study
self-dual solutions of the YM- and  SYM-theories \cite{GIOS2}. These
harmonic superspaces have the reduced values of even and odd dimensions,
and the corresponding analytic superfields parametrize moduli spaces of
self-dual solutions.

This talk is based on our work \cite{Zu0}, so we shall use notation
and conventions of this paper.

The time component of the $4D$ superfield connection $A_t$ becomes a
new $3D$-scalar superfield in the static $3D$ reduction of the $N=2$
superfield SYM-equations. We consider the $3D$-superfield BPS-type
relation between $A_t$ and the superfield strengths $W$ and $\bar{W}$
which is equivalent to the 2nd order differential constraint for
HSS-connection $\Vm$. It is shown that this 2nd order constraint
generates the standard 4th order constraint for the same connection
and all solutions of the superfield BPS-equation satisfy
the SYM-equation.

 We discuss  relations between  conjugated spinor coordinates in the
static limit of the $(3+1)$ harmonic superspace and  corresponding
pseudoreal spinor coordinates of  the 3D-Euclidean harmonic superspace.
The $N=2$ superfield BPS-condition is equivalent to
the 3D-Euclidean self-duality condition.

 Consider the nonrelativistic representation of  coordinates in
the $N{=}2,~$ $D{=}(3,1)$ harmonic superspace based on the static group
$SO(3)$
\be
x^\adb_\A\rightarrow iy^\alpha_\beta+\delta^\alpha_\beta
 t_\A~,\q \bar\theta^{\da\pm}~\rightarrow~\bar\theta^\pm_\alpha~.
\ee

Note that changes of the position of dotted $SL(2,C)$ indices after the
static reduction is connected with a convention of the conjugation of
the $SU(2)$-spinors which transforms upper indices to lower ones. The
time reduction $(t=0)$ transforms 4-dimensional harmonic superfields to
the 3-dimensional Euclidean  superfields which are covariant with respect
to the reduced supersymmetry with 8 supercharges.

Let us consider harmonic superfields $\Vp, \Vm$ of the $N=2, D=4$
SYM-theory \cite{GIK1} in the static limit
\be
\Dp \Vm-\Dm\Vp+[\Vp,\Vm]=0\lb{B3}
\ee
and construct the corresponding gauge superfields  $A_t$, $W$ and
$\bar{W}$
\bea
&&A_t={i\over2}(\bar{D}^{+}\DP) \Vm~,
\q\delta_\lambda A_t=[\lambda, A_t]~,\lb{C4}\\
&& \bar{W}=-(\DP)^2\Vm~,\q
W=(\bDP)^2\Vm~.
\eea
Superfield $A_t$  becomes covariant with
respect to residual gauge static transformations.

It seems useful to construct the manifestly supersymmetric
generalization of the Bogomolnyi equation for the $N=2$ gauge theory
in order to analyze the symmetry properties of the corresponding
monopole solutions. In addition to the $t$-reduction of  HSS-equation
\p{B3}, we propose to consider the following superfield BPS-type
relation:
\be
-2iA_t+(W+\bar{W})=-{1\over2}(D^{\alpha+}-\bar{D}^{\alpha+})
(D_\alpha^+ -\bar{D}_\alpha^+)\Vm=0~.\lb{C6}
\ee
which can be interpreted as a linear constraint on $\Vm$. Note that this
condition breaks  the $R$-symmetry $W\rightarrow e^{i\rho}W$.

The trivial  covariantly constant Abelian solutions for $A_t$ and
$W$ introduces central charges and preserves all 8 supercharges
(see e.g. \cite{IKZ}).

It is convenient to introduce the  new pseudoreal spinor coordinates
of the static Euclidean harmonic superspace
\bea
&\Theta^{\alpha\pm}\equiv{1\over\sqrt{2}}(\theta^{\alpha\pm}+
\bar\theta^{\alpha\pm})~,\q\hT^{\alpha\pm}\equiv{1\over\sqrt{2}}
(\theta^{\alpha\pm}-\bar\theta^{\alpha\pm})~,&
\lb{rot3}\\
&(\Theta^{\alpha\pm})^\dagger =\Theta_\alpha^\pm~,\q
(\hT^{\alpha\pm})^\dagger =-\hT_\alpha^\pm~.&
\eea

The corresponding 3D-Euclidean  spinor derivatives are
\bea
&&\cD^\pm_\alpha\equiv {1\over\sqrt{2}}(D^\pm_\alpha-\bar{D}^\pm_\alpha
)~,\q(\cD^\pm_\alpha)^\dagger=-\cD^{\alpha\pm}~,\lb{rot1}\\
&&\hcD^\pm_\alpha=-{1\over\sqrt{2}}(D^\pm_\alpha+\bar{D}^\pm_\alpha)~,\q
(\hcD^\pm_\alpha)^\dagger=\hcD^{\alpha\pm}~.
\eea

 It is clear that the transformation \p{rot3} connects the equivalent
$3D$ subspaces of the Minkowski and Euclidean types of harmonic
superspaces. Thus, the static equation \p{C6} is equivalent to the $3D$
limit of the Euclidean $N=2$ self-duality equation
\be
(\cD^+)^2\Vm=0~.
\ee

The component static self-dual solutions can be obtained in the
following gauge:
\bea
&&\cV^\pp={1\over2}\hT^{\alpha+}\hT^+_\alpha c(x)+{i\over2}
\Theta^{\alpha+}\hT^+_\alpha a_t(x)+i\Theta^{\alpha+}\hT^{\beta+}
a_{\alpha\beta}(x)\nn\\
&&-\hT^{\alpha+}\hT^+_\alpha \Theta^{\beta+}u_k^-\Psi^k_\beta(x)~,
\eea
where all fields are Hermitian
\be
(c, a_t, a_{\alpha\beta}, \Psi^k_\alpha)^\dagger=(c, a_t,
a^{\alpha\beta}, \Psi_k^\alpha)~.
\ee

One can derive the component $3D$ self-dual equations
\bea
&&F_{\alpha\beta}=-\nabla_{\alpha\beta} a_t~,\\
&&\nabla^{\alpha\beta}\nabla_{\alpha\beta}\,
c=2[a_t,[a_t,c]]-\{\Psi^\alpha_k,\Psi^k_\alpha\}~,\\
&&\nabla_{\beta\gamma}\bg\Psi^{\gamma k}+{i\over2}[a_t,\Psi^k_\beta]=0~,
\eea
where
\be
\nabla_{\alpha\beta}\equiv\partial_{\alpha\beta}-i[a_{\alpha\beta},]
~,\q
F_{\alpha\beta}=\partial_{\alpha\rho} a^\rho_\beta+\partial_{\beta\rho}
 a^\rho_\alpha-i[a_{\alpha\rho},a_\beta^\rho].
\ee
Note that the static gauge field strength and field $a_t$ are connected
by the self-dual Bogomolnyi equation.

The superfield analysis of the static self-duality equations
can be made by analogy with the analysis of Euclidean $4D$ self-dual
equations. The static limit of the bridge representation is
\be
A_\alpha^\pm=0~,\q\hat{A}_\alpha^+(v)=e^{-v}\hcD_\alpha^+e^v~,
\lb{vrepst}
\ee
where $v$ is the static self-dual bridge $\cD^+_\alpha v=0$.

By analogy with the 4D self-duality equation one can use
the formal identification of all groups $SU(2)$ in the $3D$ Euclidean
harmonic superspace
\be
y^{\alpha\beta}\rightarrow y^{ik},\q\Theta^\alpha_l\rightarrow
\Theta^k_l,\q\hT^\alpha_l\rightarrow \hT^k_l~.
\ee

Let us consider harmonic projections of the anti-chiral
superspace coordinates
\bea
&&y^{\A\B}_r=u^\A_ku^\B_ly^{kl}_r~,\\
&&\Theta^{\B\A}=u^{k\A} u_l^\B
\Theta^l_k~,\q\hT^{\B\A}=u^{l\A} u^\B_k\hT^k_l~,
\eea
where $A, B=\pm $.

The covariant derivative $\nabla^l_k$ are flat in the anti-chiral
self-dual representation. The corresponding bridge representation
of the $3D$ self-dual covariant derivatives has the following form:
\be
h^{-1}\frac{\partial}{\partial y^{--}_r}h~,\qq
h^{-1}\frac{\partial}{\partial \hT^{-\A}}h~,
\ee
where the anti-chiral matrix $h$ is used.

The prepotential $v^\pp=h\Dp h^{-1}$ for these solutions depends on the
(1+2) analytic coordinates $y^\pp_r$ and $\hT^+_\A$ only. It is clear
that these solutions can be interpreted as the dimensionally reduced
self-dual $4D$ solutions.

This work is supported in part by grants RFBR 99-02-18417,
INTAS-2000-254 and NATO PST.CLG 974874.

\end{document}